\documentclass[showpacs,amssymb,amsmath,nobibnotes, aps,prl,secnumarabic
,twocolumn%
]{revtex4}

\usepackage{bm}
\usepackage{natbib}
\usepackage{graphicx}

\begin{document}
\textfloatsep 10pt

\title{Phase randomization of three-wave interactions in capillary waves}

\author{H. Punzmann}
\email{Horst.Punzmann@anu.edu.au}
\author{M.G. Shats}
\author{H. Xia}

\affiliation{Research School of
Physics and Engineering, The Australian National
University, Canberra ACT 0200, Australia}

\date{\today}

\begin{abstract}
We present new experimental results on the transition from the coherent-phase to the random-phase three-wave interactions in capillary waves under parametric excitation. Above the excitation threshold, coherent wave harmonics  spectrally broaden. An increase in the pumping amplitude increases spectral widths of wave harmonics and eventually causes a strong decrease in the degree of the three-wave phase coupling. The results point to the modulation instability of capillary waves, which leads to breaking of continuous waves into ensembles of short-lived wavelets or envelope solitons, as the reason for the phase randomization of three-wave interactions.
\end{abstract}


\maketitle

Understanding instabilities of waves and nonlinear phenomena which lead to the formation of turbulence is important in a variety of nonlinear wave systems in plasma, fluids, solids, nonlinear optics, etc.
Capillary waves, which belong to a short-wave branch of the surface waves ($<10$ mm), play an important role in the surface wave physics in the ocean. They are also of interest from a general point of view as an example of nonlinear strongly dispersive waves. Capillary waves can interact via a three-wave interaction process which is possible due to the decay-type dispersion relation, $\omega_k=(\sigma/\rho)^{1/2}k^{3/2}$, where $\sigma$ is the surface tension coefficient and $\rho$ is the fluid density. Such a dispersion allows three-wave resonant conditions for the frequencies and for the wave numbers to be satisfied simultaneously: $\omega=\omega_1+\omega_2$ and $k=k_1+k_2$.

In the laboratory, capillary waves can be excited parametrically in a vertically shaken container \cite{Wright1996,Henry2000,Lommer2002,van_de_Water2009}, using electric fields \cite{Rodishevskii_1988}, ultrasonic excitation \cite{Holt 1996}, wave paddles \cite{Falcon2009}, plungers \cite{Brazhnikov2002} and other techniques. In some of these experiments, observations of fully developed broadband spectra are reported \cite{Holt 1996,Falcon2009}, while in others (e.g. \cite{van_de_Water2009}) discrete spectra are observed. In several experiments both discrete and continuous spectra are found depending on the strength of the drive \cite{Wright1996,Henry2000,Brazhnikov2002}. The observed complex wave fields are often referred to as turbulence and the shapes of the spectra are compared with the predictions of weak turbulence theory of capillary waves \cite{Zakharov&Filonenko67,Pushkarev&Zakharov2000}. It should be noted that weak turbulence theory assumes that interacting waves are weakly nonlinear and that their phases are almost random \cite{Zakharov_Lvov_Falkovich}. However, the latter assumption has never been tested in experiments. It is also not clear if the three-wave processes lead to the spreading of spectral energy into broadband turbulence in such systems.
A large body of work has also focused on the pattern formation in capillary waves (e.g. \cite{Douady1990,Gluckman1995} and \cite{Cross 1993} for review), a topic which is not addressed here.

In this Letter we focus on the physics of the formation of turbulence driven by parametrically excited monochromatic capillary waves. We show that the generation of continuous broadband frequency spectra occurs via nonlinear spectral broadening of discrete harmonics. Such broadening results from the modulation instability of capillary waves, a four-wave process of sideband generation. This nonlinear broadening changes the nature of the three-wave interactions between harmonics from coherent to the random-phase. The phenomenology of the nonlinear harmonics broadening is similar to those observed during parametric excitation of the second sound waves in superfluid helium He$^4$ \cite{Rinberg1997,Rinberg2001} and in spin waves excited in ferrites \cite{Krutsenko1978}.

The experiments are performed in vertically shaken cylindrical containers (100-200 mm in diameter, 30-50 mm deep). Above a certain acceleration threshold capillary waves on the surface of the liquid (distilled water) are generated via parametric excitation of the Faraday waves (\cite{Rodishevskii_1988,Muller1997} and also \cite{Miles_1990} for a review). The surface perturbations are detected using the reflection of a laser beam off the water surface, similarly to the method described in \cite{Brazhnikov2002}. The laser beam of 5 mm diameter is reflected onto a diffusive screen which is imaged into a photo-multiplier tube (PMT) and digitized at 100 kHz sampling rate. The reflected light intensity is related to the tilt angle $\nabla\eta$ which is the gradient of the surface elevation $\eta(r,t)$. The time-varying signal of the PMT is proportional to the fluctuating part of the reflected beam power $p(t)$. In the reported experiments, a monochromatic excitation in the range of the shaker frequencies of $f_0=$ 40 to 3500 Hz is used. This leads to the generation of the main parametrically driven $f_0/2$ subharmonic wave and a large number of its harmonics $f_n=nf_0/2$.

\begin{figure}
\includegraphics[width=8.5 cm]{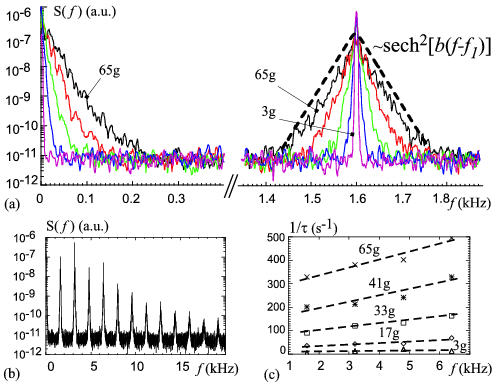}
\caption{\label{fig1} (a) Fourier power spectra of the capillary waves at different accelerations $\Delta A =3g, 17g, 33g, 41g, 65g$. The first harmonic of the parametrically excited wave at $f_1=f_0/2=1.6$ kHz and the zero-frequency sideband are shown. Dotted line corresponds to a sech$^2$-fit to the exponential spectrum. (b) The same spectrum in the range $f=(0-20)$kHz at $\Delta A =65g$. (c) Spectral width $\Delta f \propto 1/\tau$ as a function of the harmonic frequency for different accelerations.}
\end{figure}

Nonlinear broadening of capillary waves is studied under high-frequency wave excitation at $f_0=3.2$ kHz. High excitation frequency ensures that the frequency gaps between adjacent harmonics are substantially larger than the harmonic spectral widths and no overlapping occurs. Figure~\ref{fig1} shows the power spectra $S(f)$ of the signal $p(t)$. Spectra are obtained at different vertical accelerations $A$ in the range $\Delta A = 3$ to $70g$, where $\Delta A = A - A_{thr}$ denotes acceleration above the threshold of parametric instability. Close to the threshold, at $\Delta A=3g$, a narrow peak at $f_1=1.6$ kHz is observed. This wave generates over 10 harmonics, as seen in Figure~\ref{fig1}(b). As the acceleration increases, all harmonics broaden, developing exponential tails in the power spectrum which can be approximated by the squared hyperbolic secant function $\text{sech}^{2}[b(f-f_n)]$. Here, $S(f)=F_f F^*_f$ is the Fourier transform of $p(t)$ and $*$ denotes a complex conjugate. As seen in Fig.~\ref{fig1}(a) exponential tails in the spectra extend over four orders of magnitude in $S(f)$. The inverse Fourier transform of $F_f=\text{sech}[b(f-f_n)]$ is $s(t)=(\pi/b)\text{sech}(\pi^2/b \medspace t)e^{if_nt}$ which, in the time domain, describes a harmonic wave of the frequency $f_n$ modulated by a $sech$-envelope of width $\tau = b/\pi^2$. The spectral broadening $\Delta f \propto 1/\tau$ of the wave harmonic $f_n$ increases approximately linear with the increase in $\Delta A$. The increased input energy leads to spectral broadening, but does not increase the harmonic amplitudes, similar to other parametric systems \cite{Rinberg1997, Krutsenko1978}. The spectral width also increases as a function of the harmonic number at fixed $\Delta A$, as shown in Fig.~\ref{fig1}(c). As the wave spectra broaden, a zero frequency sideband develops, Fig.~\ref{fig1}(a), whose width also increases with the increase in the drive.

\begin{figure}
\includegraphics[width=8 cm]{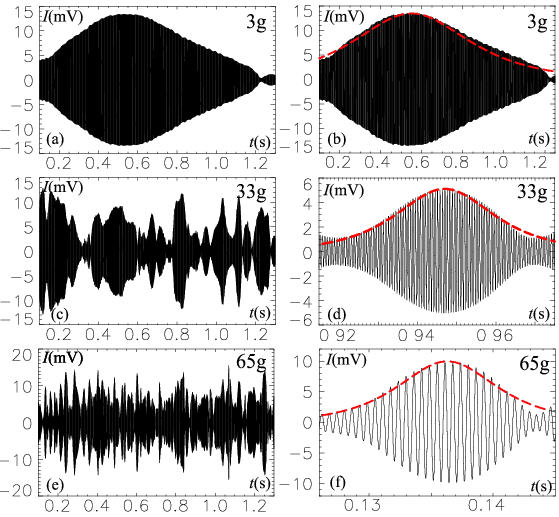}
\caption{\label{fig2} Waveforms of the bandpass-filtered signals ($f=1.6\pm 0.2$ kHz) of the first (sub)harmonic measured at different acceleration levels: (a,b) $\Delta A=3g$ ; (c,d) $\Delta A=33g$;(e,f) $\Delta A=65g$. Dashed lines in (b,d,f) correspond to the sech$(t)$ fit of the wavelet envelopes.}
\end{figure}

The development of the $f = 0$ sideband indicates that the waves are amplitude modulated. Fig.~\ref{fig2} shows the waveforms of the band-pass filtered signals of the spectrally broadened first subharmonic, $f = (1.6 \pm 0.2)$ kHz corresponding to the spectra of Fig.~\ref{fig1}. The amplitude modulation of the waves is present even very close to the threshold of parametric excitation. As the acceleration is increased, the modulation increases showing bursts of different amplitudes and durations, Fig.~\ref{fig2}(a,c,e). The shape of these modulation envelopes is well approximated by a hyperbolic secant, $s(t)=\text{sech}(\pi^2/b \medspace t)$. This fit is illustrated as a dashed line in the right column of Fig.~\ref{fig2}(b,d,f) which shows zoomed-in waveforms of the corresponding signals from the left column. The modulated wave appears as a sequence of the $sech$-modulated wavelets of various amplitudes and widths. Thus, the frequency-domain and the time-domain representations are consistent with the amplitude modulation of the wave harmonics.

The observed amplitude modulation of capillary waves is likely to result from the modulation instability whose linear theory was presented over 40 years ago \cite{Zakharov1968} but has not been studied in any detail, in contrast to the modulation instability of the surface gravity waves, or the Benjamin-Feir instability \cite{Benjamin&Feir67}. The Lighthill criterion of the modulation instability, $(\partial \omega/\partial|a|^2)(\partial^2 \omega_k/\partial k^2)<0$, is satisfied for capillary waves as well as for gravity waves. Here, $a$ is the wave amplitude and $\omega= \omega_k [1-(ka)^2/16]$ includes a nonlinear frequency correction \cite{Zakharov1968}. A well-known exact solution of the nonlinear Schr\"odinger equation which describes the evolution of the modulationally unstable waves \cite{Zakharov1968} is the hyperbolic secant envelope soliton, found in many physical systems including the gravity surface waves \cite{Yuen_Lake75}. The above observation of the $sech$-modulated wavelets points to a similar phenomenon in capillary waves.

The nonlinear spectral broadening of parametrically excited waves described above has also been reported in the second sound waves in liquid helium He$^4$ \cite{Rinberg1997} and in spin waves excited in ferrites \cite{Krutsenko1978}. In both systems, above the threshold of the parametric excitation, waves develop exponential frequency spectra. Similarly to our observations, it was found that the spectral width and wave amplitude modulation frequency of the second sound wave increases with the drive \cite{Rinberg1997} (the amplitude of the first sound wave in their case), similar to our result of Fig.~\ref{fig2}. Theoretical interpretation of the spectral broadening in these experiments is given in terms of the four-wave scattering \cite{Krutsenko1978,Lvov_Cherepanov1978}, or a two-step three-wave interaction process \cite{Muratov1997}.

\begin{figure}
\includegraphics[width=8 cm]{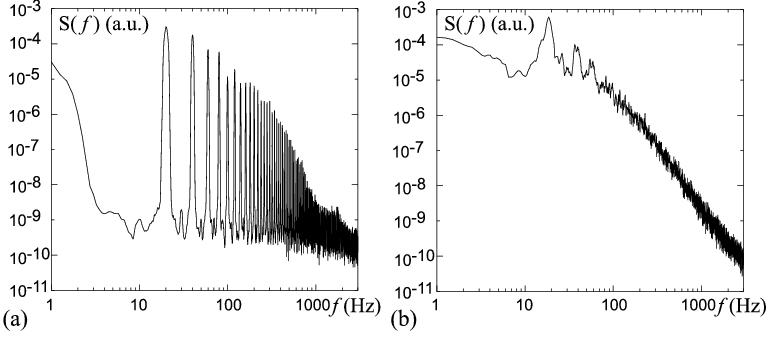}
\caption{\label{fig3} (a) Discrete and (b) continuous spectra of capillary waves driven by the parametrically excited wave at $f_1=f_0/2=20$ Hz. The r.m.s. acceleration is (a) $\Delta A=2.5g$ and (b) $\Delta A=15g$.}
\end{figure}

Next we show results on the low frequency wave excitation. In this case, the spectral broadening of the wave harmonics can lead to their overlap. Figure~\ref{fig3}(a) shows a $\nabla\eta$ power spectrum of the waves excited at $f_0=40$ Hz, $f_1=20$ Hz at the acceleration of $\Delta A=2.5g$. More than $n=50$ harmonics are observed  in this case which extend over two decades in $f$. A spectrally broadened zero-frequency sideband is also seen. When the acceleration is increased to $\Delta A=15g$ a continuous broadband spectrum is observed, Fig.~\ref{fig3}(b). Higher acceleration, beyond $\Delta A=15g$, leads to further steepening of the surface ripple and to the formation of water droplets.

As mentioned above, the dispersion relation of capillary waves allows three-wave interactions. To characterize a degree of the phase coupling in discrete and continuous spectra shown in Fig.~\ref{fig3}, we compute the auto-bicoherence, or normalized bispectrum \cite{Kim79}. The auto-bispectrum of the reflected laser power $p(t)$ is defined as $B(f_1,f_2)=\left\langle{F_{f} F^*_{f_1} F^*_{f_2}}\right\rangle=\left\langle{A_{f_1} A_{f_2} A_{f} e^{\phi_f-\phi_{f_1}-\phi_{f_2}}}\right\rangle$, where $f=f_1+f_2$. Here angular brackets denote the ensemble averaging. If the phases of waves at $f_1, f_2$ and $f$ are statistically random, the average value of the bispectrum is zero. The auto-bicoherence is the squared auto-bispectrum normalized by the auto power spectra of the interacting waves:
\begin{equation}\label{Eq:Bic}
b^2_f(f_1,f_2)=\frac{\left|\left\langle{F_{f} F^*_{f_1} F^*_{f_2}}\right\rangle \right|^2}{\left\langle{F_{f}F^*_{f}}\right\rangle \left\langle{F_{f_1}F^*_{f_1}}\right\rangle \left\langle{F_{f_2}F^*_{f_2}}\right\rangle}.
\end{equation}
\noindent It changes between 0 (no phase coupling) and 1 (coherent waves) and reflects the strength of the three-wave interactions. We also compute the summed bicoherence, $SB\left( {f} \right) =
{\sum\limits_{f = f_{1} + f_{2}}  {b_{f}^{2} (f_{1} ,f_{2} )}} $. This quantity
gives a measure of the total phase coupling to the frequency $f$ from all frequencies $f_{1}$ and $f_{2}$ in the spectrum satisfying $f = f_{1} + f_{2} $.

\begin{figure}
\includegraphics[width=8 cm]{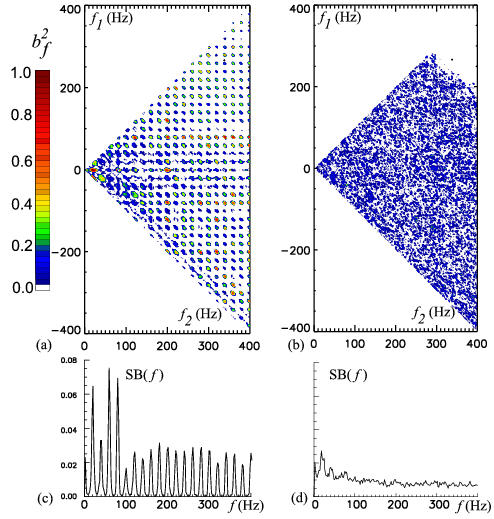}
\caption{\label{fig4} (a,b) The auto-bicoherence and (c,d) the summed bicoherence of the reflected light intensity computed for the conditions of (a,c) discrete spectrum of Fig.~\ref{fig3}(a), and (b,d) continuous spectrum of Fig.~\ref{fig3}(b).}
\end{figure}

Figure~\ref{fig4} shows the auto-bicoherence $b^2_f$ and the summed bicoherence $SB\left( {f} \right)$ corresponding to the spectra in Fig.~\ref{fig3}. The degree of the phase coupling between the discrete harmonics of Fig.~\ref{fig3}(a) is very high with $b^2_f=(0.5-0.9)$ as shown in Fig.~\ref{fig4}(a). This is indicative of coherent phase interactions. In the broadband turbulence case of Fig.~\ref{fig3}(b), the level of the bicoherence drops below 0.2 at all frequencies, Fig.~\ref{fig4}(b), even at the strongest harmonics of the pumping wave. However, the bicoherence of the continuum increases, as seen in the calculated summed bicoherence by comparing figures~\ref{fig4}(c) and ~\ref{fig4}(d). This suggests that at the increased excitation amplitude a large number of waves participate in three-wave interactions which have almost random phases.

\begin{figure}
\includegraphics[width=8 cm]{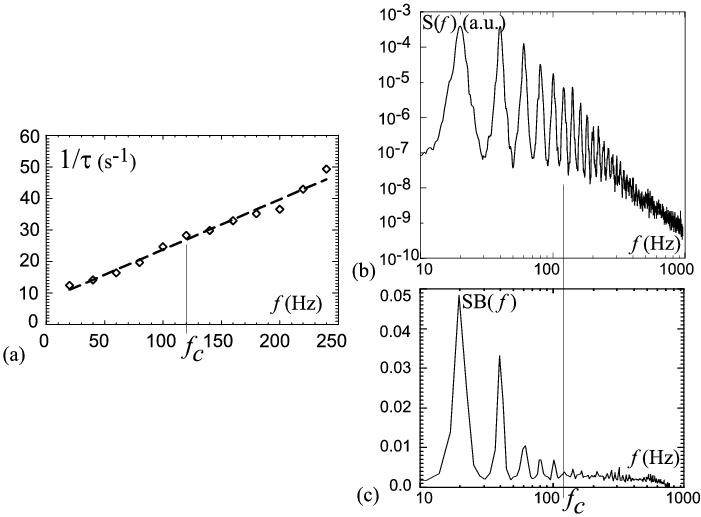}
\caption{\label{fig5} (a) Spectral width of the wave harmonics as a function of their frequency, corresponding to the spectrum of Fig.~\ref{fig5}(b). (b) Power spectrum and (c) summed bicoherence of the reflected light intensity of capillary waves driven by the parametrically excited wave at $f_1=f_0/2=20$ Hz at intermediate acceleration of $\Delta A=6g$.}
\end{figure}

Similarly to the case of high frequency excitation, the wave harmonics at low frequency excitation show $sech$-spectra. Their spectral width also increases with the increase in acceleration and with the harmonic number, Fig.~\ref{fig5}(a). The increase in the spectral width results from the decrease in the average width $\tau$ of the wavelet envelopes.

As the acceleration is gradually increased, higher harmonics start overlapping, such that the continuous spectrum is first generated at higher frequencies. This is shown in the spectrum of Fig.~\ref{fig5}(b) obtained at the intermediate acceleration of $\Delta A=6g$. In this spectrum the wave continuum is observed in the frequency range $f > 350$ Hz. However the wave harmonics in the range $f > f_c \backsimeq 120$ Hz already lose their coherence, Fig.~\ref{fig5}(c). The summed bicoherence is reduced to the random phase level in the range $f > f_c$. The data of Fig.~\ref{fig5} (dotted line) suggest that the transition from coherent to random phases occurs at $f_1 \tau \lesssim 1$, where $f_1=20$ Hz is the frequency of the first parametric subharmonic and $\tau \approx 0.04$ s is derived from the harmonic spectral width at $f = f_c$. This can be explained as follows. The degree of coherence in three-wave interactions depends on the nonlinear interaction time between waves in the wave triads. Thus it is defined by the time width $\tau$ of the shortest-lived wave (in our system the highest frequency) in the resonant triad. When this width becomes shorter than the period $T=1/f_1$ of the lowest frequency harmonic, at $\tau/T < 1$, the bicoherence is reduced.

The experiments show that gradual development of the wave continuum occurs due to the spectral broadening of harmonics. Whether or not the Kolmogorov-type cascade via 3-wave interactions \cite{Zakharov&Filonenko67,Pushkarev&Zakharov2000} transfers energy within the spectrum cannot be decided solely based on the spectrum shape, as is often done. To further investigate this, one would need to analyze the wave kinetic equation using experimentally measured wave fields, similar to the method described in \cite{Ritz1986}.

Summarizing, we have shown that parametrically excited capillary waves on the water surface become spectrally broadened. The frequency spectra of the wave harmonics show exponential tails and are well approximated by the sech$(bf)$-fit. The increase in nonlinear broadening with the increase in the drive, is linked to the development of the modulation instability. In the time domain, the instability leads to breaking of continuous waves into $sech$-modulated wavelets or envelope solitons. A decrease in the average width of the wavelet envelopes eventually leads to the reduction in the nonlinear interaction time within resonant triads and to the drop in the degree of phase coupling between three waves.

\begin{acknowledgments}
We are grateful to V. Lebedev, S. Lukaschuk, G. Falkovich and V. Steinberg for useful comments. This work was supported by the Australian Research Council's Discovery Projects funding scheme (DP0881571).
\end{acknowledgments}

\end{document}